%% file: engmikheevaetal.tex
\documentclass[
aps,%
12pt,%
final,%
notitlepage,%
oneside,%
onecolumn,%
nobibnotes,%
nofootinbib,%
superscriptaddress,%
noshowpacs,%
centertags]%
{revtex4}
\usepackage{graphicx}

\begin{document}

\title{CATALOG OF SUPERMASSIVE BLACK HOLES FOR INTERFEROMETRIC OBSERVATIONS}

\author{\firstname{E.~V.}~\surname{Mikheeva}}
\email{helen@asc.rssi.ru}
\author{\firstname{V.~N.}~\surname{Lukash}}
\email{lukash@asc.rssi.ru}
\author{\firstname{S.~V.}~\surname{Repin}}
\email{sergerepin1@gmail.com}
\author{\firstname{A.~M.}~\surname{Malinovsky}}
\email{ingirami@gmail.com}
\affiliation{%
P.~N.~Lebedev Physical Institute of Russian Academy of Sciences
}%

\begin{abstract}
   The paper presents a catalog of supermassive black holes (SMBH) shapened for the 
interferometric observations in millimeter and submillimeter wavelength ranges and 
based on the open sources. The catalog includes the name of the object, coordinates, 
angular distance, the mass, the angular size of the gravitational radius of SMBH, 
the integral flux of radiosource, related  with SMBH, in the range $20\div 900$~GHz 
which is considered to be used in Event Horizon Telescope, future space mission 
Millimetron and others.

      The catalog is intended for use during the planning of the interferometric observations
of SMBH shadows.

\end{abstract}

\maketitle

\section{Introduction}

      Black holes are the most interesting physical objects in the Universe. The possibility of
the existence of  bodies whose  gravitational field is so strong that even the light cannot escape
their gravitational ``pit''  was first considered by J.~Mitchell in 1783. In 1796 the
similar reasoning was expressed  by P.-S.~Laplace.  However, until the creation of the
relativistic theory of gravity, the idea of the existence of ``dark stars'' remained
purely speculative. After the appearance of the General Theory of Relativity (GR) the
situation has changed. The exact solution of the Einstein equations for a point-like mass was
first obtained by K.~Schwarzschild at the turn of 1915 and 1916. In the spring of 1916
the same solution was presented in the thesis of O.~Droste  (whose scientific supervisor
was H.A.~Lorenz) in the form, which later became standard. Soon after that in 1918
H.~Reissner and G.~Nordstr\"{o}m found the solution of the Einstein equations for a massive
body with an electric charge. Many years later, the solutions of the General Relativity
equations for a massive rotating body (Kerr metric)  and a rotating charged body
(Kerr-Newman metric) were also found.

       Along with the search of the exact solution of General Relativity equations the physicists
tried to comprehend what are the properties of a body described by the Schwarzschild metric
(Kerr metric, etc.) or simply answer the question: what this object is? Following the successful
joke of the famous witty  J.A.~Wheeler this body in 1968 was assigned a name of a ``black
hole''\footnote{It seems the author of this name is Ann Ising who first used it as early as
in 1964; see \cite{petrov2015}, p.~152.}. For the objects described by the Schwarzschild
and Kerr metrics, the corresponding names were fixed: Schwarzschild black hole,  Kerr black
hole, and so on. A lot of interesting facts on the history of black holes can be found in in the popular 
science book (\cite{petrov2015}), two essays \cite{levin1} and \cite{levin2},  
and in the review \cite{zakharov1999}.

      In 1935 A.~Einstein and N.~Rosen considered a structure that was the union of two black
holes ``sewed together'' exactly at the gravitational radius (Einstein--Rosen bridge).
The purpose of the creation of this construction was to avoid the problem of singularity, i.e.
the appearance of the configurations with infinite curvature.  Later, the family of black holes
acquired the ``relatives'': white holes (\cite{Novikov1965}), wormholes
(\cite{MTY1988}) and  black-and-white holes (see \cite{lukash2012}, \cite{strokovlukashmikheeva2016}).

      The aim of this paper is not to investigate what is hidden beneath the horizon of black
holes, but to present for interferometric observations a compiled catalog of known black holes
located in the central regions of galaxies.  The reason to accent the attention on supermassive
black holes (with  a mass $M \ge 10^4 M_\odot$)  is that they are the most convenient objects
for observations because their angular size is greater than the one for the stellar mass black holes.

     It seems there are four ways to create a black hole. In the first case black hole may be born at 
a final stage of a massive star evolution. Such a star can belong to any generation of stars. Also, 
black hole can appear as a result of gas instability that result is direct collapse. The third way is 
related with star-dynamical processes in star clusters. More details concerning these three ways 
can found in \cite{Volonteri2010}. Finally, black hole may appear in result of collapse of an 
overdensity region. In this case we deal with primordial black hole (PBH).

     It is widely known that the black hole emits nothing  to surrounding space  except the
Hawking radiation, that has the quantum nature. So, it is desirable to make clear the sense of the
expression ``the black hole observations''.  Throughout the decades, this meant the
radiation of matter in the gravitational field of a black hole. More specifically, this meant the
radiation generated inside or near the jet and/or accretion disk.  At present time the ``black
hole observation'' includes also the investigation of a black hole shadow or silhouette.

      The form of a black hole shadow is determined by several parameters. Two of them refer
to a black hole. In the frameworks of GR they are the mass of a black hole and its momentum.
The others describe the source of the photons emission. If it is the accretion disk the important
parameters are the dependence of the emission on the radial coordinate and the angle between
the normal to the disk plane and the view line of a distant observer. Nevertheless, for a wide range
of the parameters the shadow can be considered as a circle and its diameter is approximated by
the expression (see \cite{diametershadow}, \cite{brightring}):
\begin{equation}
         \theta_{shadow} \sim 10.4\, \frac{r_g}{D_A},
         \label{gravrad}
\end{equation}
where $r_g$ is the gravitational radius (here $r_g = GM/c^2$), $D_A$~-- the angular distance to
the black hole. Substituting the parameter values of the stellar black holes and the black holes 
presented in the central part of the distant galaxies and comparing $\theta_{shadow}$, one can 
conclude that  the SMBH shadows are more reliable objects for the observations than stellar black holes. 
For the black hole in the center of our Galaxy the estimation of the angular size of its shadow gives 
$\theta_{ring} \approx 53$~microsecond arc.

       The angular distance depends on the parameters of the cosmological model and reaches the 
maximum $D_A \approx 1750$~Mpc (at $z \approx 1.6$), for the Hubble constant  
$H_0=70$~km/(s$\cdot$Mpc), the cosmological matter density $\Omega_m=0.3$ and the 
$\Lambda$-term density 0.7. Having set the angular resolution of Millimetron (in interferometric mode) 
as an angular size of the shadow one can derive from (\ref{gravrad}) that Millimetron will detect all SMBHs  
with mass $M>10^9 M_\odot$  in the Universe if they are bright enough.  The black hole of such mass 
can be called ``hypermassive''. The mission will also detect the nearby SMBHs with lower masses. 

     Despite the absence of published observations of black hole shadows, they
are already planned to be used for testing of the General Relativity in strong gravitational fields (see,
for example \cite{KBHSH}).
The theorem on the ``absence of hair'' will be apparently the key point in the verification
of General Relativity. The experts associate the special hopes with observations in millimeter and
submillimeter bands (\cite{diametershadow, april2018}). The observations in the submillimeter
range have the privilege because the influence of the scattering processes in interstellar and
intergalactic medium decreases with increasing the frequency, see \cite{refraction}. This is extremely important for the
image reconstruction procedure in the interferometric observations.

\section{Catalog structure}

       By now there are several SMBH catalogs. The main ones are  (in order of publication):
\cite{Gultekin2009}, \cite{McConnel2012}, \cite{Saglia2016} and \cite{vanderBosch2016}.

     The content and format of the catalogs were determined by the tasks and requirements of
researchers. The original purpose of creating our catalog was to select the SMBH, the shadows of
which could be resolved by space observatory Millimetron. This imposes the restrictions not only
on the angular size of SMBH, but also on their celestial coordinates and fluxes in the frequency
channels of Millimetron. The table with 20 best candidates for observations presented in
\cite{ivanovufn2018}.

      It is planned that the Millimeterron Observatory will operate in wide frequency bands centered 
at 22, 43, 100, 240 and 350~GHz and, optionally, at 600 and 800 GHz (\cite{kardashev2014}). 
As we work with the published data, it turned out that for a large number
of SMBHs the total emission fluxes in these bands are unknown. For some sources, the nearest
measured points are spaced by several orders of magnitude in frequency. For this reason, and also
because of a possible change of the Millimetron orbit, we decided that it is necessary to compile as
complete catalog of SMBH as possible.

      At present, the observations of black hole shadows are also planned for other telescopes.
The most promising project is the Event Horizon Telescope (EHT)\footnote{EHT --
Event Horizon Telescope, {\tt http://eventhorizontelescope.org}}, which is the array of several
observatories operating in the interferometer mode. Due to the larger area of the receiving antennas,
the sensitivity of such an interferometer is better than that of the Millimetron, but its angular resolution
is fundamentally limited by  the Earth diameter. Given this limitation, the list of shadows available for 
the observation consists of only a few objects. First of all, it is a black hole in the center of the Galaxy, 
known as the source of Sgr~A$^*$ and the active nucleus of M87 galaxy. The expected angular sizes
of these shadows differ by a factor of two; they are approximately 50 and 20 microseconds arc.

      The proposed catalog of SMBH includes  353 objects.  This is much more than the preliminary list
of 20 sources mentioned above. This is due to both the small angular size of
most of the known SMBH, and the absence of any information on flows from these objects in
the submillimeter range.

     The catalog is available in website: 
{\tt http://millimetron.ru/index.php/en/scientific-} {\tt program/the-catalog-of-supermassive-black-holes}. 
It  is organized as follows.

\subsection{Source name}

      The first column shows the source name.  Sometimes the object has one or more synonyms, which
are also presented. This can help find the object in the astronomical databases.  As one can see from
the names of the SMBH, some of them are the nuclei of nearby galaxies and some are active galactic
nuclei (QSOs and Seyfert galaxies).

\subsection{SMBH mass}

     The second column shows the mass of the black hole in the units of $10^8 M_\odot$. We call this
quantity as a ``mass parameter''.

      As it follows from the table, the range of object masses is rather large. The most massive single
SMBH  has the mass parameter of more than 100, while the lower mass boundary of the SMBH, by
definition, is not less than $10^{-4}$. For a number of black holes, there are several measurements
of mass, which differ by a factor of 5. The statistical error differs from object to object and for the
majority of SMBH is about 30\%. For individual objects the uncertainty of the mass measurement could
reach a few order of magnitude. For 15\% of the SMBHs, only the upper estimation of the mass is
available. 

     Today there are several ways to determine the black hole mass: the dynamics of stars, gas
dynamics, masers, reverberation method (\cite{brightring}) and a number of statistical methods.
Another method was proposed in \cite{brightring}, in terms of the angular size of the black
hole shadow. For obvious reason this method has never been used yet.

\begin{figure}[!htbp]
\renewcommand{\baselinestretch}{1}
\centerline{\includegraphics[width=130mm]{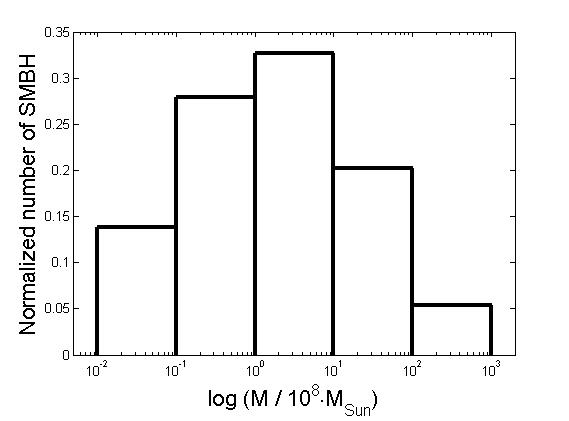}}
  \caption{Percentage of SMBH as function of the mass parameter (see text).}
  \label{Distribution_1}
\end{figure}

     The histogram in Fig.~\ref{Distribution_1} represents the distribution of  SMBH  mass. The 
normalized number of SMBHs is shown here as a function of a ``mass parameter''.

\subsection{Distance to the object}

     The third column of the catalog shows the distance to the object, expressed in megaparsecs.

     As it is known a few types of distance are used in cosmology. They are: the metric distance,
the distance based on luminosity, the same based on the angular scale and the distance based on
the redshift of the object.  In SMBH catalogs the type of the distance representation is not usually
indicated, but the metric distance is  traditionally used.

     Our catalog is intended to the observation of the black hole shadows. So, we prefer the angular 
distance because it is the one which we use to evaluate the angular scale of the shadow. It is easy 
to evaluate the distance according to the angular scale if we know the redshift of the emission source 
and accept the cosmological model. The following set of parameters has been used: the Hubble 
constant $H_0 = 70\,\, \text{km/(s}\cdot \text{Mpc)}$, the cosmological matter density 
$\Omega_m = 0.3$, the $\Lambda$-term density $\Omega_\Lambda = 0.7$. These values of the 
parameters differ somewhat from so-called ``post-Planckian''  cosmological parameters, 
but using of the latter would be excessively accurate.

\begin{figure}[!htbp]
\renewcommand{\baselinestretch}{1}
\centerline{\includegraphics[width=130mm]{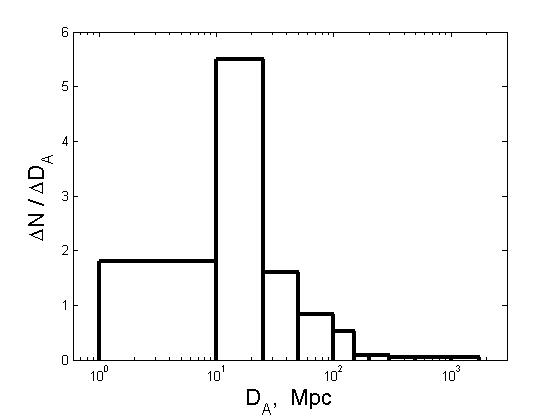}}
  \caption{Averaged number of SMBH in spherical layer as function of its angular distance.}
  \label{Distribution_2}
\end{figure}

     The histogram in  Fig.~\ref{Distribution_2} demonstrates the averaged number of SMBH in spherical 
layer as function of its angular distance $D_A$.  The distance step on this histogram is not 
constant (1-10, 10-25, 25-50, 50-100, 100-150, 150-200, 200-300, 300-500, 500-1000 and 
1000-1750~Mpc) and this may cause some inconvenience. 

\subsection{Angular size of SMBH}

\begin{figure}[!htbp]
\renewcommand{\baselinestretch}{1}
\centerline{\includegraphics[width=130mm]{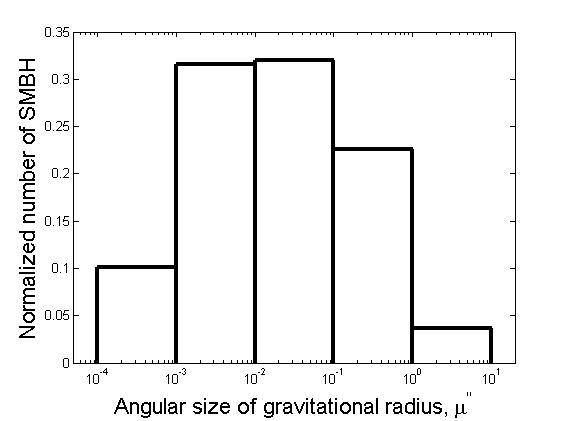}}
  \caption{Percentage of SMBH as function of its angular size of gravitational radius.}
  \label{Distribution_3}
\end{figure}

     The fourth column indicates the angle at which the gravitational radius of the black hole should 
be observed, expressed in arc microseconds. We emphasize that this is the gravitational radius in
the context of expression (\ref{gravrad}) and the angular size of the shadow is about 10 times larger.  
Fig.~\ref{Distribution_3} shows the percentage of SMBH as function of its angular size of gravitational 
radius.  

\subsection{Object coordinates}

     In the fifth and sixth columns, the coordinates of the SMBHs are given in the equatorial 
coordinate system for the J2000 epoch. For most sources we limited the accuracy of coordinates 
to one second arc. This is more than enough to determine the projection of the base when planning 
the interferometric observations of the shadows.

\subsection{Frequency}

     The seventh column shows the central frequencies of the channels (expressed in gigahertz) in which 
the SMBHs were observed. Since our catalog is ``sharpened'' for the interferometric 
observations of shadows in the millimeter and submillimeter ranges, where no significant self-absorption 
of the radiation is expected, we tried to work at the frequency range of $22-800$~GHz. However, for some 
sources, there were few observation points, in this case we cataloged the data with near frequencies. For 
example, the source PG~0052+251 has not  currently observational data within the interval $22-800$~GHz, so 
we provide data for the near frequencies, 9 and 1764~GHz.

\subsection{Total flux from SMBH}

     The eighth column contains the data of the magnitude of radiation fluxes at the corresponding frequencies.
The flux values are expressed in Jansky. The main, but not the only source of information on the fluxes was 
the public database NED\footnote{http://ned.ipac.caltech.edu/}.  If a number of observations (at different time 
and with different receiving equipment) were carried out for the frequency channel of interest, we indicated the 
range of the observed quantities. It should be noted that these fluxes characterize not the SMBH itself, but cover 
the larger area than the size of shadow. The reason for this is obvious: until now a few objects have been 
observed in the required frequency range with a really high angular resolution,  comparable with the size of the 
gravitational radius of SMBH. Therefore, the actual value of the flux will, of course, be less, than indicated in 
the Catalog.

\subsection{Comment}

     The last column of the table contains the reference to the source catalog (catalogs) and/or to 
the published article.

\section{Subcatalogs}

      Two more subcatalogs for future interferometric observations presented in Tables \ref{Table_2}  
and \ref{Table_3}.  The first of them (Table~\ref{Table_2}) summarizes the top 25 black holes ordered 
descending by  their mass parameter.  One more selection presented in Table~\ref{Table_3}, where 
summarized the top 25 SMBHs with highest angular size.  Some objects (NGC~4889, NGC~1600) are included 
in both subcatalogs.

\section{Conclusion}

      As it follows from the available data, the observation of any representative image of a black hole 
shadow is a very difficult task: on the one hand it is necessary that the source appears bright enough 
at the frequency of the intended observations, on the other hand, its angular scale should be several 
times greater than the angular resolution of the interferometer at the same frequency. 

      In this situation, it is encouraging that for a large number of sources the fluxes in the submillimeter 
range remain unknown yet and some of them may appear bright enough.

     The catalog of supermassive black holes, which we present here is intended for use in the planning 
of the interferometric observations of SMDM shadows in the submillimeter and millimeter ranges.
It is available in website of Astro Space Center of P.~N.~Lebedev Physical Institute of Russian Academy 
of Sciences:
{\tt http://millimetron.ru/index.php/en/scientific-program/the-catalog-of -supermassive-black-holes}. 

      This study was supported by the Russian Foundation for Basic Research, grant 16-02-01043,
Program of Fundamental Research by the Russian Academy of Science Presidium P7 (subprogram
Transitional and Explosive Processes in Astrophysics) and a state order for scientific program 
``Millimetron''. Authors thank P.B.~Ivanov (LPI, Moscow) for fruitful discussions. One author 
(S.R.) is very grateful to Dr. R.Beresneva, Dr. O.Suvenkova  and Dr. O.Kosareva for the possibility of 
fruitful working on this problem.

\input{engtableofsmbh.tex}

\end{document}

%% file: engtableofsmbh.tex
\begin{table*}[!h]
\caption{Top 25  supermassive black holes in the Catalog}
\label{Table_2}
\bigskip
\begin{tabular}{|c|c|c|c|c|c|c|c|c|}
            \hline
Object      & $M$,                     & $D_A$,    &  Angular          & Right             &   Decli-   & Fre-          & Flux,   &     \\
                 & $10^8M_\odot$   &    Mpc       &    size,              & ascension,  & nation,   &   quency,   &    Jy      &   Reference   \\
                 &                               &                  & $\mu$sec arc   &  h. m. s.       &  deg.      &   GHz         &           &  \\
            \hline
 1              &            2                 &    3          &      4           &   5               &     6              &    7            &   8        &    9       \\
            \hline
S5                   & 400     & 1531    & 0.26   & 00 17 08.5   & + 81 35 08  &  15       & 0.47-0.916 & \cite{Ghisellini2010} \\
0014+81        &             &              &            &                       &                     &   30-31 & 0.63-1.4     &   \\
\hline
H1821            & 300  & 912 (z) & 0.32  & 18 21 57.3   & + 64 20 36   & 93        & 0.01 & \cite{Walker2014} \\     
+643              &          &              &           &                       &                       & 1800   & 1 &  \\       
                         \hline
APM                & 230 & 1447 (z) & 0.16 & 08 31 41.7 & + 52 45 18 & 100 & 0.001 & \cite{Riechers2009}  \\
08279+5255  & 100 &  1447(z) & 0.07 &                     &                    & 250  & 0.034 & \cite{Saturni2016}  \\
IRAS F            &     &         &   &            &            &300 & 0.06&   \\
08279+5255  &     &         &   &            &            &660 & 0.34&   \\
                        &     &         &   &            &            &857 & 0.386  & \\
 \hline             
NGC 4889       & 210 & 103   & 2.01    & 13 00 08.13 & + 27 58 37.2 & 2.4       & 0.001     & \cite{McConnel2012} \\       
                          & 209 & 102  &  2.02    &                       &                        & 3000   & $<0.069$   &  \cite{Saglia2016},  \cite{vanderBosch2016}\\   
  \hline
SPT-CL          & 200 & 1374 & 0.15 & 23 44 42.2 & - 42 43 08 & 1000 & 0.02 & \cite{McDonald2012} \\    
J2344-4243  &  &  &  &  &  &  &  & \\       
         \hline
SDSS             & 195 & 1554 (z) & 0.13 & 07 45 21.78 & + 73 43 36.1 &  & in NED & \cite{Zuo2015} \\
J074521.78   &        &                 &          &                       &                        &  & absent &  \\
+734336.1     &  &  &  &  &  &  & &  \\
			\hline
OJ 287       & 180 & 930  & 0.19    & 08 54 48.88 & + 20 06 39.6 & 22-23 & 2.4-6.1 & \cite{Valtonen2012}  \\ 
                   &   1   & 930  & 0.001  &                       &                        & 43 & 1.6-2.9 & secondary BH \\
                   &        &           &            &                        &                        & 100 & 4.5 & \\     
                   &        &          &             &                        &                        & 240 &1.4-3 & \\
                   &        &          &             &                       &                        & 340 &4.5 &  \\
                   &        &          &             &                       &                        & 800 & 1.3-1.7 & \\
                    \hline
 NGC 1600     & 170 & 64 & 2.6   & 04 31 39.9 & - 05 05 10.0 & 5        & 0.016 &  \cite{vanderBosch2016} \\  
                        &        &       &         &                     &                       & 3000 & 0.190  &  \\
                       \hline
SDSS            & 151 & 1513 (z) & 0.10 & 08 08 19.69 & + 37 30 47.3 & 14000 & 0.002  & \cite{Zuo2015} \\
J080819.69  &         &                &          &                       &                       &              &             &  \\
+373047.3   &          &                &          &                       &                       &              &             &  \\
[1mm]
  \hline
\end{tabular}
\end{table*}

\addtocounter{table}{-1}
\begin{table*}[!h]
\caption{Top 25  supermassive black holes in the Catalog \textit{(continued)}}
\bigskip
\begin{tabular}{|c|c|c|c|c|c|c|c|c|}
            \hline
 1              &            2                 &    3          &      4           &   5               &     6              &    7            &   8        &    9       \\
			\hline
SDSS              & 141 & 1522 (z) & 0.09 & 11 59 54.33 & + 20 19 21.1 &325000  & 0.0006  & \cite{Zuo2015} \\
J115954.33     &        &  &      &             &              &  &   &  \\
+201921.1       &  &  &  &  &  &  & &  \\
                     \hline
SDSS           & 135 & 1470 (z)  & 0.09 & 08 04 30.56 & + 54 20 41.1 & 14000 & 0.004  & \cite{Zuo2015} \\
J080430.56 &         &                 &          &             &              &  &  & \\
+542041.1   &  &  &  &  &  &  & &  \\
                       \hline
SDSS            & 124   & 1146 (z)   & 0.10    & 01 00 13.02     & - 28 02 25.8    &            &   in  NED  & \cite{Wu2015}\\
J0100            &           &                  &             &                          &                          &            &  absent      & \\
+2802            &          &                   &             &                          &                          &            &   & \\
                       \hline
SDSS       & 123 & 1496 (z) & 0.08 & 07 53 03.34 & + 42 31 30.8 & 30  & 0.058 & \cite{Zuo2015} \\
J075303.34 &     &                 &      &             &              &  & &  \\
+423130.8  &  &  &  &  &  &  & &  \\
			\hline
SDSS       & 120 & 1479 (z) & 0.08 & 08 18 55.77 & + 09 58 48.0 & 14000 & 0.007  & \cite{Zuo2015} \\
J081855.77 &     & &      &             &              &  &  &  \\
+095848.0  &  &  &  &  &  &  & &  \\
         \hline
SDSS       & 112 & 1508 (z) & 0.07 & 08 25 35.19 & + 51 27 06.3 & 14000 & 0.004 & \cite{Zuo2015} \\
J082535.19 &     & & &             &              &  &  &  \\
+512706.3  &     &  &  &  &  &  & &  \\
                         \hline
SDSS             & 110 & 1273 (z) & 0.09 & 01 31 27.34 & - 03 21 00.1 &  & in  NED & \cite{Ghisellini2015} \\
J013127.34   &        &                &          &                       &                       &  & absent & \\
-032100.1       &  &        &    &             &              &  &       & \\
                    \hline
Holmberg       & 100      & 224 (z)  & 0.45 & 00 41 50.5  & - 09 18 11 & 22.5 & 0.002     & \cite{LopezCruz2014}\\
15A                 & (10-     &               &           &                     &                   & 43     & $<0.002 $ & \\
*MCG-02-       & 3100) &               &           &                     &                   &          &                & \\
02-086            &            &               &           &                      &                   &          &                & \\
                          \hline
RX J1532.9   & 100 & 1009.4 (z) & 0.10 & 15 32 53.8 & + 30 20 58 &$3\times10^8$  & $2\times 10^{-6}$ & 
\cite{Hlavachek2013} \\       
+3021          &     &  &      & &  &  &  &  \\            
                        \hline
PKS 2126 & 100 & 1546 & 0.06 & 21 29 12.2 & - 15 38 41 & 20-24   & 1.07-0.84   & \cite{Ghisellini2010} \\       
-158           &         &           &          &                     &                    & 41-43  & 0.6-0.5   &  \\       
                   &        &            &           &                    &                     & 90-94 & 0.3-0.5          &  \\       
                   &       &              &           &                   &                     & 230     & 0.08              &  \\       
                   &       &             &            &                  &                      & 312     & $<0.8$               &  \\ 
[1mm]
  \hline
\end{tabular}
\end{table*}

\addtocounter{table}{-1}
\begin{table*}[!h]
\caption{Top 25  supermassive black holes in the Catalog \textit{(continued)}}
\bigskip
\begin{tabular}{|c|c|c|c|c|c|c|c|c|}
            \hline
 1              &            2                 &    3          &      4           &   5               &     6              &    7            &   8        &    9       \\                                      
                     \hline
PSO                 & 100 & 1720 (z)   & 0.06 & 22 16 48.6 & + 01 24 27 &  & in NED &  \cite{Liu2015}, dbl. BH.  \\       
O334.2028      &        &                   &         &                     &                     &  & absent      &  orb. per.  542 days \\   
+01.4075         &  &  &  &  &  &  &  &   \\    
                       \hline
SDSS              & 98 & 1499  (z)  & 0.07 & 01 57 41.57 & - 01 06 29.6 &  & in  NED & \cite{Zuo2015}\\
J015741.57    &      &                   &         &             &              &  & absent & \\
-010629.6       &    &         &      &             &              &  &       & \\
                         \hline
NGC 3842      & 97   & 98.4  & 0.97  & 11  44  02.15 & + 19 56 59.3 & 2.4     & 0.022 & \cite{McConnel2012} \\
                         & 91   & 92.2 & 0.97   &                        &                         & 3000 & 1.49   & \cite{Saglia2016}, \cite{vanderBosch2016}                            \\
                          \hline
SDSS                & 91 &1511 (z)  & 0.06 & 23 03 01.45 & - 09 39 30.7 & 325000 & 0.0004  & \cite{Zuo2015} \\    
J230301.45      &    &                &      &             &              &  &     &  \\       
-093930.7         &  &  &  &  &  &  &  &  \\
                       \hline  
NGC 5419    & 72   & 56.2 & 1.27   & 14 03 38.7  & - 33 58 42  & 5           & 0.09-0.12 & \cite{Saglia2016},  \cite{vanderBosch2016} \\      
     		       &        &          &           &                      &                     & 1900    & $<0.021$ & 
     		       \\   
                           \hline
CID-947            & 69 &1537 (z) & 0.04 & 10 01 11.35 & + 02 08 55.6 & 100 & 0.0001& \cite{Ghisellini2010} \\
                           &       &               &          &                       &                        & 300 & 0.003 & \\
[1mm]
  \hline
\end{tabular}
\end{table*}

\begin{table*}[!h]
\caption{Top 25  supermassive black holes ranged on angular size}
\label{Table_3}
\bigskip
\begin{tabular}{|c|c|c|c|c|c|c|c|c|}
            \hline
Object           & $M$,                          & $D_A$, &  Angular          & Right         &   Decli-      & Fre-      & Flux,   &     \\
                 & $10^8M_\odot$   &    Mpc       &    size,     & ascension,   & nation,   &   quency,       &    Jy      &   Reference   \\
                 &                               &               & $\mu$sec arc     &  h. m. s.       &  deg.       &   GHz  &  &  \\
                 &                 &           &             &      &        &         &         &              \\
            \hline
 1              &            2                 &    3          &      4           &   5               &     6              &    7            &   8        &    9       \\
			\hline  
Srg A*   & 0.041     & 0.008     & 5.06 & 17 45 40.02 & - 29 00 28.17 & 43    & 1.3-1.9     & \cite{Gultekin2009}, \cite{McConnel2012}   \\       
              & 0.0431   & 0.00833 & 5.11 &                      &                          & 100  & 2.1-2.4     &  \cite{Saglia2016}, \cite{vanderBosch2016}    \\   
              &                              &                &         &                       &                         & 240  & 2.8-4.1    &   
               \\
             &                              &                &          &                       &                         & 340   & 3              &   
               \\    
                     \hline  
NGC 4486 & 36 & 17       & 2.09    & 12 30 49.42 & + 12 23 28.0 & 22-23   & 0.5-21     &  \cite{Gultekin2009}\\       
M 87           & 63 & 17       & 3.66    &                      &                         & 41         & 3.6-13.5  & \cite{McConnel2012} \\   
                    & 62 & 16.7    & 3.67   &                      &                          & 94-100 &0.5-5.3    & \cite{Saglia2016}, \cite{vanderBosch2016} \\       
                    &      &             &            &                      &                          & 300       & 1.3 &  \\         
                    &      &             &           &                       &                          & 600       & 1.4 &  \\   
                   &      &              &           &                       &                          & 860       & 1 &  \\       
			\hline 
NGC 4649     & 21 & 16.5 & 1.23    & 12 43 40.4 & + 11 33 10 & 10.5     & 0.018 &   \cite{Gultekin2009} \\       
M 60               & 47 & 16.5  & 2.81   &                     &                    & 1700    & $<0.1$   &    \cite{McConnel2012}, \cite{Saglia2016},  \cite{vanderBosch2016} \\      
                        \hline
NGC 1600     & 170 & 64 & 2.6   & 04 31 39.9 & - 05 05 10.0 & 5        & 0.016 &  \cite{vanderBosch2016} \\  
                        &        &       &         &                     &                       & 3000 & 0.190  &  \\
                           \hline
NGC 4889       & 210 & 103   & 2.01    & 13 00 08.13 & + 27 58 37.2 & 2.4       & 0.001     & \cite{McConnel2012} \\       
                          & 209 & 102  &  2.02    &                       &                        & 3000   & $<0.069$   &  
                          \cite{Saglia2016}, \cite{vanderBosch2016} \\     
                          \hline
NGC 224      & 1.5        & 0.8    & 1.85     & 00 42 44.35 & + 41 16 08.6   & 5         & 0.036     & \cite{Gultekin2009},  \cite{McConnel2012} \\
M 31              & 1.4        & 0.77   & 1.80   &                        &                         & 1900  & 7800      &  \cite{Saglia2016}, \cite{vanderBosch2016}  \\
                         \hline
NGC 1407      & 45   & 28  & 1.6  & 03 40 11.8 & -18 34 48 & 5         & 0.034  & \cite{Saglia2016}, \cite{vanderBosch2016}\\
                        &         &        &        &                    &                   & 1900  & 0.092  &  \\
                          \hline
NGC 4472       & 25   &17.1   &1.44   & 12  29  46.7    & +  08  00  02& 15   & 0.004 & \cite{Saglia2016}, \cite{vanderBosch2016} \\       
M 49                 &        &           &           &                         &                       & 96   &  0.15   &    \\
                         &        &           &           &                         &                       &1667& $<0.09$  &        \\       
                        \hline
NGC 3706      & 59  & 46 & 1.3 & 11  29  44.4 & - 36  23  29 & 5 & 0.025  & \cite{vanderBosch2016}\\
                         &       &       &       &                      &                      & 1900 & $<0.022$ &   \\
                        \hline
NGC 3923    & 28  & 20.9  & 1.3   & 11 51 01.7   & - 28  48  22 & 5       & 0.001 & \cite{Saglia2016}, \cite{vanderBosch2016}\\
                        &       &           &         &                       &                     & 1900 & 0.048 &  \\
[1mm]
  \hline
\end{tabular}
\end{table*}

\addtocounter{table}{-1}
\begin{table*}[!h]
\caption{Top 25  supermassive black holes ranged on angular size \textit{(continued)}}
\bigskip
\begin{tabular}{|c|c|c|c|c|c|c|c|c|}
            \hline
 1              &            2                 &    3          &      4           &   5               &     6              &    7            &   8        &    9       \\                                                                       
                       \hline  
NGC 5419    & 72   & 56.2 & 1.27   & 14 03 38.7  & - 33 58 42  & 5           & 0.09-0.12 & \cite{Saglia2016}, \cite{vanderBosch2016} \\       
     		       &        &          &           &                      &                     & 1900    & $<0.021$ &  \\   
			\hline 
NGC 3842      & 97   & 98.4  & 0.97  & 11  44  02.15 & + 19 56 59.3 & 2.4     & 0.022 & \cite{McConnel2012} \\
                         & 91   & 92.2 & 0.97   &                        &                         & 3000 & 1.49   & \cite{Saglia2016}, \cite{vanderBosch2016} \\
                       \hline
NGC 5055     & 8.3 & 8.7 & 0.94 & 13  15  49.3 & + 42  01  45 & 15   & $<0.001$ &  \cite{vanderBosch2016}\\
                        &       &        &         &                       &                       & 300 & 1.3         &  \\
                        &       &        &         &                       &                       & 600 & 2.6        &  \\
                        &       &        &         &                       &                       & 850 & 64         &  \\
                         \hline
NGC 3115        & 9.6    & 10.2 &   0.93  & 10 05 14.0     & - 07 43 06.9   & 1.4   & 0.0006  &  \cite{Gultekin2009}, \cite{McConnel2012}  \\
                          & 9.0    & 9.5   &   0.94   &                        &                        & 1900 & $<0.045$ & \cite{Saglia2016}        \\
                          & 8.8    &  9.5  &   0.91   &                        &                        &           &              &  \cite{vanderBosch2016} \\ 
	\hline
IC 1459           &   28      & 30.9 & 0.89 & 22 57 10.61 & - 36  27  44.0  &    20   &  0.55        & 
\cite{Gultekin2009}, \cite{McConnel2012} \\    
                        &   25       & 28.9 & 0.85 &                       &                          &    95   &  0.26  &  
                        \cite{Saglia2016}, \cite{vanderBosch2016} \\               
                        &              &          &          &                       &                          & 1800 &  1.1-2.4     & \\
                      
  \hline
NGC 4374     & 15    & 17.0  & 0.87  & 12 25 03.74 & + 12 53 13.14 & 15        & 0.16-1.3     & \cite{Gultekin2009} \\
M 84               & 8.5  & 17.0   & 0.49  &                       &                          & 43        & 0.1             & 
\cite{McConnel2012} \\
                        & 9.2 & 18.5    & 0.49  &                       &                         & 95-100 & 0.14-0.17  & \cite{Saglia2016} \\ 
                        & 9.3 & 18.5   & 0.50  &                        &                         & 350      & 0.15            & 
                        \cite{vanderBosch2016} \\
                        &       &            &           &                        &                        & 670      & 0.12             &  \\
                        \hline
NGC 1550     & 37   & 51.6   & 0.72  & 04  19  37.9 & + 02 24 34  & 2.3        & 0.008   & \cite{Saglia2016}; \\
                        &        &              &            &                     &                      & 3000  & $<0.245$ & 
                        \cite{vanderBosch2016}  \\
                          \hline
NGC 5328      & 47   & 64.1   & 0.72  & 13 52 53.3 & - 28 29 22 & 5       & $<0.0009$ & \cite{Saglia2016}, \cite{vanderBosch2016} \\       
                         &        &            &           &                    &                    & 3000 & $<0.07$     & 
                          \\   
                         \hline
NGC 6861     &   20   &  27.3    &  0.72   & 20 07 19.5   & - 48  22  13   & 0.8     & 0.015 & \cite{Saglia2016}, \cite{vanderBosch2016} \\       
                       &           &              &             &                      &                        & 3000  & 3-3.5 & 
                        \\  
                           \hline
NGC 3091        &   36   & 51.3   & 0.70 & 10 00 14.3 & - 19 38 13 & 5            & 0.007 & \cite{Saglia2016}, \cite{vanderBosch2016} \\
                           &          &           &         &                    &                    & 12500     & 0.003 & 
                            \\
                         \hline
NGC 4594      & 5.7 & 10.3   & 0.55 & 12 39 59.43 & -11 37 23.0 & 20  & 0.08             & \cite{Gultekin2009}  \\       
M 104             & 5.3 & 10.3    & 0.51 &                      &                      & 250  & 0.19-0.44  &\cite{McConnel2012}  \\   
Sombrero     & 6.7 & 9.9      & 0.67 &                      &                      & 350 & 0.24-0.92   & \cite{Saglia2016}  \\       
                       & 6.6 & 9.9      & 0.66  &                      &                      & 600 & 5.6              & 
                       \cite{vanderBosch_2016} \\   
                         &       &            &            &                      &                     & 850 & 12.1             &  \\   
[1mm]
  \hline
\end{tabular}
\end{table*}

\addtocounter{table}{-1}
\begin{table*}[!h]
\caption{Top 25  supermassive black holes ranged on angular size \textit{(continued)}}
\bigskip
\begin{tabular}{|c|c|c|c|c|c|c|c|c|}
            \hline
 1              &            2                 &    3          &      4           &   5               &     6              &    7            &   8        &    9       \\                                                                       
                      \hline
NGC 5128 & 3           & 4.4      & 0.67  & 13 25 27.61 & - 43 10 08.8 & 22           & 3-112    & \cite{Gultekin2009}, \cite{McConnel2012}\\        
Cen A         & 0.7        & 4.4     & 0.16    &                      &                       & 41           & 32-72   & \cite{Gultekin2009}, \cite{McConnel2012}
  \\  
                    & 0.57     &  3.62  &  0.16   &                      &                       & 90-93      & 41        & \cite{Saglia2016},  \cite{vanderBosch2016} \\   
                    &              &            &            &                       &                       & 230-240  & 5.8-6  & 
                    \\       
                   &               &            &           &                        &                       & 350          & 18       &  \\      
                        \hline
NGC 1277    & 47 & 71 & 0.65  & 03 19 51.49 & + 41 34 24.7 & 3000  & $<0.7$     & \cite{vanderBosch2016} \\  
                       \hline
NGC 1332    & 14.5 & 22.3 & 0.64     & 03  26  17.321 & - 21  20  07.33 & 5        & 0.005 & \cite{McConnel2012},  \cite{Saglia2016}\\
                       & 6.8   & 22.3 & 0.30     &                           &                           & 1700  & 1.56   &  \cite{vanderBosch2016}\\
                       &          &         &             &                           &                           &             &         &  
                      \\
                          \hline  
NGC 1399      & 5.1        & 21.1    & 0.24   & 03  38  29.08   & - 35  27  02.67 & 8.5    & 0.36 & \cite{Gultekin2009},  \cite{McConnel2012}   \\
                         & 13         & 21.1    & 0.61   &                          &                           & 1875 & 0.02 & \cite{Gultekin2009},  \cite{McConnel2012}
                         \\
                         & 8.8        & 20.9    & 0.42   &                          &                           &           &          & \cite{Saglia2016}\\
                         & 8.7        & 20.9    & 0.41   &                          &                           &           &          & 
                         \cite{vanderBosch2016}\\
[1mm]
  \hline
\end{tabular}
\end{table*}